\documentclass[11pt]{article}
\usepackage{graphicx}

\newcommand{\BABARPubYear}    {01}

\newcommand{\BABARProcNumber} {67}
\newcommand{\SLACPubNumber} {9026}

\input ./babarsym

\def\Dztokpi    {\ensuremath{\Dz \to K^{-}\pi^{+}}\xspace}
\def\Dzbtokpi   {\ensuremath{\Dzb \to K^{+}\pi^{-}}\xspace}
\def\DztokpiWS  {\ensuremath{\Dz \to K^{+}\pi^{-}}\xspace}

\def\mKpi       {\ensuremath{m_{K\pi}}\xspace}

\def\dm         {\ensuremath{\Delta m}\xspace}

\newcommand{\kevcc}{\ensuremath{{\mathrm{\,Ke\kern -0.1em V\!/}c^2}}\xspace}

\def\Rws        {\ensuremath{R_{W\!S}}\xspace}

\long\def\inst#1{\par\nobreak\kern 4pt\nobreak
    {\it #1}\par\vskip 10pt plus 3pt minus 3pt}

\begin{document}
{\pagestyle{empty}

\begin{flushright}
SLAC-PUB-\SLACPubNumber \\
\babar-PROC-\BABARPubYear/\BABARProcNumber \\
October, 2001 \\
\end{flushright}

\par\vskip 4cm

\begin{center}
\Large \bf Determination of the wrong sign decay rate
  $\mathbf{\DztokpiWS}$ and the sensitivity to $\mathbf{\Dz-\Dzb}$
  mixing.
\end{center}
\bigskip

\begin{center}
\large 
Ulrik Egede \\
Blackett Laboratory, Imperial College, \\
London SW7 2BW, United Kingdom.\\
(for the \lbabar\ Collaboration)
\end{center}
\bigskip \bigskip

\begin{center}
\large \bf Abstract
\end{center}
The \Dz meson can decay to the wrong sign \Kp\pim state either through
a doubly Cabibbo suppressed decay or via mixing to the \Dzb state
followed by the Cabibbo favoured decay \Dzbtokpi. We measure the rate
of wrong sign decays relative to the Cabibbo favoured decay to $(0.383
\pm 0.044 \pm 0.022)\%$ and give our sensitivity to a mixing signal.

\vfill
\begin{center}
Contributed to the Proceedings of the 
International Europhysics Conference on HEP, \\
12---18 July 2001, Budapest, Hungary
\end{center}

\vspace{1.0cm}
\begin{center}
{\em Stanford Linear Accelerator Center, Stanford University, 
Stanford, CA 94309} \\ \vspace{0.1cm}\hrule\vspace{0.1cm}
Work supported in part by Department of Energy contract DE-AC03-76SF00515.
\end{center}

\section{Introduction}
\label{sec:Introduction}
Particle-antiparticle mixing between neutral mesons arises when the
mass eigenstates of the production Hamiltonian are not the same as the
weak eigenstates which are responsible for the meson decay.

Due to the presence of the weak interaction the physical states are
thus a superposition of the mass eigenstates. This superposition
splits the mass of the physical states and introduces the possibility
of mixing between the mass eigenstates in the form of oscillations.
Mixing is defined in terms of two dimensionless parameters: $x =
\Delta M / \Gamma$ and   $y = \Delta \Gamma / 2\Gamma$ where $\Delta M
= m_{2} - m_{1}$ and  $\Delta \Gamma = \gamma_{2} - \gamma_{1}$ are
the differences between the masses and the decay rates of the strong
eigenstates respectively, and $\Gamma = (\gamma_{2} + \gamma_{1})/2$. A recent
review of the predictions for the level of mixing can be found
in~\cite{Nelson:1999fg}.

\section{Event Selection}
\label{sec:Selection}
The results presented in this work are based on data collected with
the \babar\ detector~\cite{babar} at the \pep2 asymmetric \epem
storage ring at the Stanford Linear Accelerator Center during the
1999--2000 Run 1. This corresponds to an integrated luminosity of
$20.6$~fb$^{-1}$ recorded \emph{on-resonance} at the \FourS mass and
2.6~fb$^{-1}$ \emph{off-resonance} about 40~\mevcc below this energy.

\Dz candidates produced in \ccbar continuum events are selected
through the decay chain $\dsp \ra \Dz \pi^{+}_{s}$ followed by the
decay $\Dz \ra K^{\pm} \pi^{\mp}$. In this way the production flavour
is tagged by the charge of the slow pion from the \dsp decay. The
decay is then classed as a \emph{right sign decay} if the Kaon has the
opposite charge of the slow pion $\pi^+_s$ and a \emph{wrong sign
  decay} if they have the same charge. The charge conjugated \dsm
decay is treated in the same way.

The event selection criteria are: the momentum of the \dsp in the \FourS rest
frame above 2.6~\gevc; particle identification of both \Dz daughters;
good track and vertex quality; helicity cut on the Kaon decay angle with
respect to the \Dz momentum evaluated in the \Dz rest frame, and $p_t >
0.5$~\gevc for the pion from the \Dz. Finally if multiple overlapping
candidates are left in an event the event is rejected. A common vertex
fit is made to the \Dz, the \dsp and the beam spot taking advantage of
the small beam spot size $(\sx,\sy,\sz) \approx (120\mum, 5.6\mum, 
7.9\mm)$.

\section{Analysis method}
\label{sec:Analysis}
An unbinned log likelihood fit is performed using the values of \mKpi,
$\dm=m_{(K\pi)\pi_s} - \mKpi$, the proper time $t$ and its estimated
error for each \dsp candidate. In these variables the \textit{right
  sign} signal has a very simple shape. It peaks in the mass
distributions and follows an exponential convoluted with our
resolution model for the time evolution. The \textit{wrong sign}
signal has, under the assumption of no \CP violation, the time
evolution modulated by the mixing parameters $x'$ and $y'$:
\begin{equation}
\label{eq:DecayRate}
   \Gamma ( \Dzb(t) \ra {K^{-}\pi^{+}} ) =  \Gamma ( \Dz(t) \ra {K^{+}\pi^{-}} )  \approx  
    e^{-t/\tau} \left[ 
      {R} + 
      \sqrt{R}{y'} t/\tau  + 
      \frac{1}{4} \left( {x'}^{2} + {y'}^{2} \right) t^{2}/\tau^2
    \right]  
\end{equation}
and convoluted with the same resolution function as the \textit{right
  sign} decay. $R$ is the time integrated doubly Cabibbo suppressed
decay rate. The parameters $x'$ and $y'$ are related to the mixing
parameters $x$ and $y$ through a rotation $(x',y') \equiv (x
\cos\delta + y \sin \delta, y \cos\delta - x \sin\delta)$ where
$\delta$ is the unknown phase difference between the Cabibbo favoured
and doubly Cabibbo suppressed decay.

In order to have a reliable measurement of the mixing rate we need a
good understanding of the background sources in the $\left( \mKpi,\dm
\right)$ plane and of their decay time evolution. The background
categories we model are: a real \Dz combined with a fake slow pion; an
incomplete \Dz like $\Dz \ra \Km\ellp\nul$ reconstructed as \Dztokpi;
reflections of $\Dz \ra \Kp\Km / \pip\pim$; swapped particle ID
hypothesis of the $K$ and the $\pi$ in the \Dz decay, and purely
combinatoric background.

\begin{figure}[htbp]
  \centering
\includegraphics[width=\linewidth]{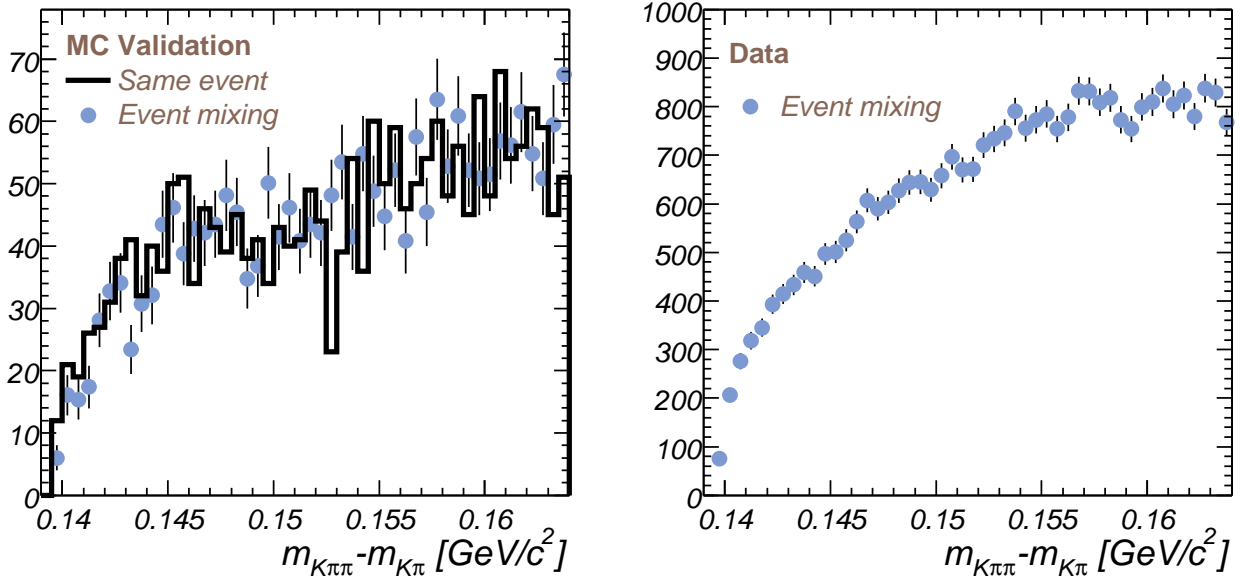}
\caption{To the left a comparison between the shape obtained from event
  mixing and the true background shape where the \Dz and slow pion are
  from the same event. To the right the \dm background shape obtained
  from event mixing on data.}
\label{fig:EvtMixTrueD0}
\end{figure}

We use event mixing as a method to obtain the \dm distribution for the
combinatorial and fake slow pion categories directly from data. The
idea is to reconstruct \dsp candidates from slow pions in one event
with \Dz candidates from other events. In this way it is assured that
a reconstructed \dsp really has a fake slow pion. In
Fig.~\ref{fig:EvtMixTrueD0} we show a validation of the method on
Monte Carlo and the actual \dm distribution we used obtained directly
from the data.

\section{Results}
\label{sec:Results}
In table~\ref{tab:sourcesFractionsFullFit} we list the fractional
contributions for signal and background sources as obtained from the
fit. In Fig.~\ref{fig:fitResult} we show the comparison between the
fit and the data.

\begin{table}[htbp]
  \centering
\begin{tabular}{lcc}
\hline
Source & Right sign (\%) & Wrong sign (\%) \\
\hline
signal                &  $92.16 \pm 0.15$  & $6.25 \pm 0.57$  \\ 
Real \Dz fake $\pi_s$ &  $4.57  \pm 0.11$  & $56.5 \pm 1.4  $ \\ 
Incomplete \Dz and reflections   &  $0.742 \pm 0.072$ & --               \\ 
Swapped \Dz           &  --                & $1.29 \pm 0.35$  \\ 
Combinatoric          &  $2.525 \pm 0.081$ & $36.0 \pm 1.1$   \\ 
\hline
\end{tabular}
\caption{Fractional contribution of signal and background sources
  obtained from the simultaneous fit to the \textit{right sign} and
  \textit{wrong sign} sample. $1.804\gevcc < \mKpi < 1.924\gevcc$, $\dm
  < m_\pi + 25\mevcc$.}
\label{tab:sourcesFractionsFullFit}
\end{table}
\begin{figure}[htbp]
  \centering
\includegraphics[width=\linewidth]{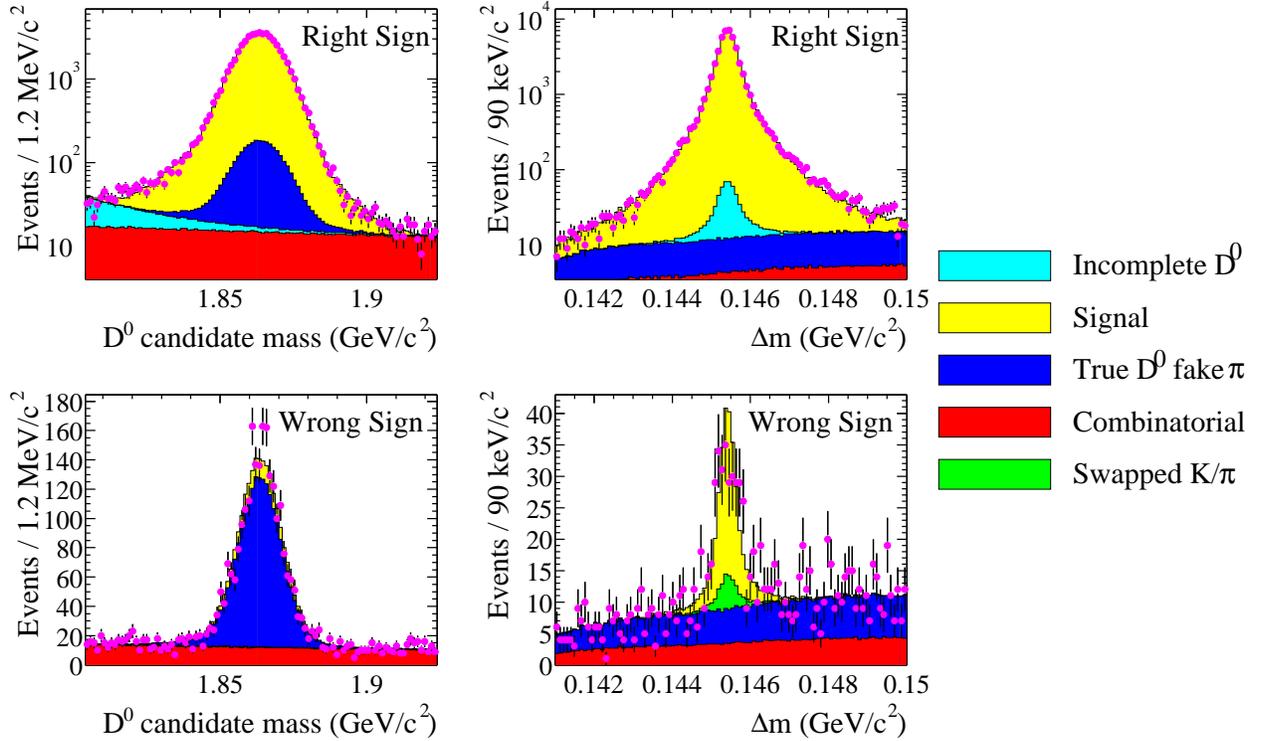}
\caption{A comparison between the data represented as points with
  errors and the overall fit to the \textit{right sign} and
  \textit{wrong sign} \dsp candidates. Notice the logarithmic scale
  for the \textit{right sign} decay.}
\label{fig:fitResult}
\end{figure}

In total the selected right sign sample has 58723 candidates and the
wrong sign sample 3315 candidates. If we combine this with the signal
fractions in table~\ref{tab:sourcesFractionsFullFit} we get 54120
right sign signal events and 210 wrong sign signal events. The ratio
between the \textit{wrong sign} signal and the Cabibbo allowed decays
is then $\Rws = (0.383 \pm 0.044 ) \%$.

The systematic checks we performed have focused on the log likelihood
fit, the selection criteria and detector effects. The mixing
parameters are strongly anti-correlated and the likelihood space
stretches to a non physical region. For this reason, when considering
the systematic checks on the mixing parameters, rather than comparing
the minimum values obtained from fits to different configurations, we
will compare the one and two sigma likelihood contours. A summary of
the systematic errors on \Rws are given in table~\ref{tab:Systematics}
and contours for the different systematic checks are overlaid in
Fig.~\ref{fig:sysContour}. The systematic effect from the internal
alignment of the silicon tracker is pending the reprocessing of the
data and the central value of the mixing fit is kept blinded until
then.

\begin{table}[htbp]
  \centering
\begin{tabular}{llc}
\hline
Type & Variation & Error (\%) \\
\hline
Kaon identification & Loose---Tight           & 0.001 \\
Pion identification & Loose---Tight           & 0.010 \\
Kaon $p_t$ cutoff   & 0.1--0.5~\gevc          & 0.009 \\
$\cos(\theta^\ast)$ & 0.65--1.0               & 0.006 \\
\Dz mass window     & $\pm40$--$\pm80$~\gevcc & 0.010 \\
\dm window          & 15--28~\mevcc           & 0.004 \\
SVT track quality   &                         & 0.011 \\
Background shape    &                         & 0.003 \\
Background fractions &                        & 0.005 \\
$p^\ast_{\dsp}$ cutoff & 1.4--2.8~\gevc       & 0.004 \\
Prob($\chi^2$) vertex fit & 0.002--0.05       & 0.001 \\
Other               &                         & 0.002 \\
\hline
Sum in quadrature   &                         & 0.022 \\
\hline
\end{tabular}
\caption{Summary of the systematic errors on \Rws.}
\label{tab:Systematics}
\end{table}

\begin{figure}[htbp]
  \centering
  \includegraphics[width=0.7\linewidth]{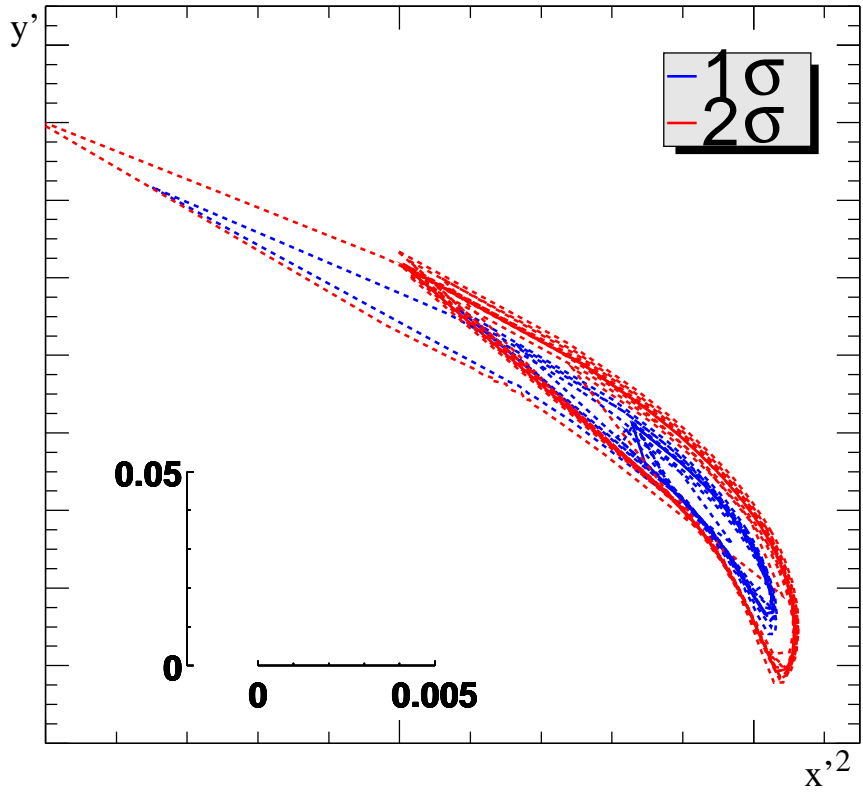}
  \caption{Superposition of all the contours for the systematic checks. The
    central value of the fit is blind.}
  \label{fig:sysContour}
\end{figure}
\begin{figure}[htbp]
  \centering
  \includegraphics[width=0.7\linewidth]{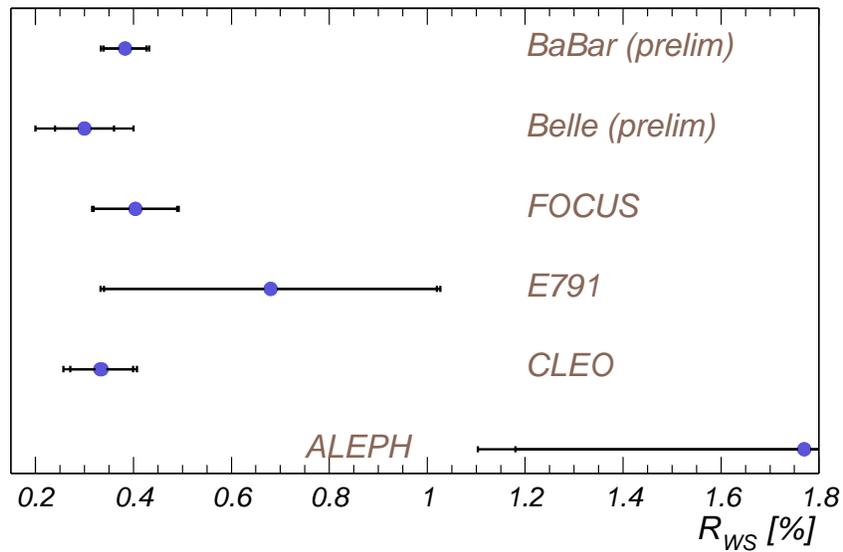}
  \caption{Experimental values for \Rws. The results from \babar\ and Belle
    are preliminary.}
  \label{fig:DCSDplot}
\end{figure}

By adding in quadrature the systematic errors  we obtain the following
preliminary result for the wrong sign signal fraction that, in the
assumption of no mixing, corresponds to the doubly Cabibbo suppressed
decay rate:
\begin{equation}
\Rws = (0.383 \pm 0.044(Stat.) \pm 0.022(Sys.) ) \%.
\end{equation}
This value is compared with other experimental
results~\cite{Ban:2001}\phantom{\cite{Godang:1999yd}}\phantom{\cite{Aitala:1998fg}}\phantom{\cite{Barate:1998uy}}
\hspace{-1.3cm}--\cite{Link:2000kr}
in Fig.~\ref{fig:DCSDplot}.

\bibliographystyle{JHEP-2}
\bibliography{proc}

\end{document}